\documentclass[twocolumn,secnumarabic,amssymb, nobibnotes, aps]{revtex4}
\usepackage{amssymb}
\usepackage{color}
\usepackage{graphicx}
\usepackage{dcolumn}
\usepackage{bm}
\usepackage[header,title,page,titletoc]{appendix}
\usepackage{amsmath}
\usepackage{subfigure}
\usepackage{mathrsfs}
\usepackage{amsfonts}
\usepackage{amssymb}
\usepackage{amsfonts}
\providecommand{\U}[1]{\protect \rule{.1in}{.1in}}

\newcommand{\D}{\mathrm{d}}
\newcommand{\p}{\partial}

\begin{document}
\title{Emergent Quantum Dynamics of Vortex Line under Linear Local Induction Approximation}
\author{Guihao Jia}
\email[Corresponding: ]{guihao.jia@mail.bnu.edu.cn}
\affiliation{Center for Advanced Quantum Studies, Department of Physics, Beijing Normal University, Beijing, 100875, China}
\author{Yu Xu}
\affiliation{Center for Advanced Quantum Studies, Department of Physics, Beijing Normal University, Beijing, 100875, China}
\author{Xiao Kong}
\affiliation{Center for Advanced Quantum Studies, Department of Physics, Beijing Normal University, Beijing, 100875, China}
\author{Cuixian Guo}
\affiliation{Center for Advanced Quantum Studies, Department of Physics, Beijing Normal University, Beijing, 100875, China}
\author{Silei Liu}
\affiliation{Center for Advanced Quantum Studies, Department of Physics, Beijing Normal University, Beijing, 100875, China}
\author{Su-Peng Kou}
\affiliation{Center for Advanced Quantum Studies, Department of Physics, Beijing Normal University, Beijing, 100875, China}

\begin{abstract}
Using the linear local induction approximation, we investigated the self-induced motion of a vortex line that corresponds to the motion of a particle in quantum mechanics. Concerning Kelvin waves, the effective Schr\"{o}dinger equation, physical quantities operators, and the corresponding path-integral formula are obtained. The vortex line-particle mapping may help in understanding particle motion in quantum mechanics.
\end{abstract}
\maketitle

\section{Introduction}

One hundred and fifty years ago, Lord Kelvin studied the properties of vortex lines, consisting of fluid rotating around a centerline. He predicted that the vortex line could carry linear waves (Kelvin waves)\cite{kelvin}, which were confirmed by experiment in $~^{4}\mathrm{He}$ superfluid\cite{ajhj} 50 years ago. In subsequent years, single vortex lines have attracted much attention by researchers. In 1967, Batchelor\cite{Batchelor1970AN} presented a formulation of a non-local three-dimensional Biot--Sarvart equation to approximate the local expression of motion, which became known as the local induction approximation (LIA). In 1972, Hasimoto\cite{1972JFM....51..477H} developed a transformation that mapped the LIA equation onto the nonlinear Schr\"{o}dinger equation (NLS), and identified the soliton-wave solutions along the vortex line. From experiments in 1998, Donnelly and Barenghi\cite{waaaa} reported on the properties of vortex lines in $~^{4}\mathrm{He}$ superfluid at low temperatures. Recently, quantum turbulence based on LIA and NLS has attracted considerable attention among researchers. The energy transferred between Kelvin waves of different wave-numbers through nonlinear coupling is believed to be the mechanism underlying superfluid turbulence\cite{Boffetta2009Modeling}. The relevant work on vortex line motion is more concentrated on soliton solutions of NLS\cite{Salman2014Multiple}\cite{Salman2013Breathers} and quantum turbulence\cite{Vinen2002}\cite{Laurie2009The}.

From another perspective, quantum mechanics (QM) is a fundamental theory of modern physics that has successfully explained most experimental results and has broad application. The Schr\"{o}dinger equation is fundamental in QM\cite{1926AnP...384..361S}. During the last 100 years, scientists have discovered several models of motion (approximately) described by the Schr\"{o}dinger equation, such as the propagation of acoustic waves in plasma\cite{333}, the motion of light pulses in nonlinear optical fibers\cite{333}, and the recently discovered evolution of self-gravitational-induced waves in disks surrounding a massive body\cite{Batygin2018Schr}. Here, we point out that under the linear LIA (LLIA), the self-induced motion of a vortex line is characterized by an effective Schr\"{o}dinger equation that makes it possible to visualize the QM wave function intuitively.

In this paper, we investigate from theory the correspondence between the self-induced motion of the vortex line and the particle motion in QM under this LLIA. In Sec.~\ref{sec1}, we present a brief derivation of the Schr\"{o}dinger-like equation from the Biot--Savart equation applying the LLIA, the redefined vortex line wave function being different from that of Hasimoto's\cite{1972JFM....51..477H}, and, defining a characteristic evolution time, the condition under which linearity is valid. In Sec.~\ref{sec2}, the momentum operator is naturally deduced from the original definition in the fluid. After considering the angular momentum operator and Hamiltonian, the effective Schr\"{o}dinger equation and the corresponding path-integral formula based on LLIA are subsequently obtained in Sec.~\ref{esepi}.

\section{Self-induced motion of vortexline under LLIA}

\label{sec1} From the original Biot--Savart equation, we derive briefly the expression of the self-induced motion of a vortex line that satisfies a Schr\"{o}dinger-like equation under the LLIA.

The velocity $\vec{v}$ associated with a vortex line is described by the Biot-Savart equation
\begin{equation}
\vec{v}=\frac{\Gamma}{4\pi}\int_{L^{\prime}}\frac{\mathrm{d}\vec{l}^{\prime}\times \vec{R}}{|\vec{R}|^{3}},
\end{equation}
where $\Gamma$ denotes the vorticity of vortex line, $L^{\prime}$ the path of integration, $\mathrm{d}\vec{l}^{\prime}$ the element of integration, and $\vec{R}$ the position vector from the source point to the field point. The Helmholtz vorticity theorem states that $\Gamma$ remains the same value along the vortex line.

Assuming that the vortex line is a single-valued function of $z^{\prime}$ in the coordinate system $x^{\prime}y^{\prime}z^{\prime}$, the self-induced motion indicates that apart from the point itself, each element on the vortex line has an inductive effect on a certain point $(x_{0},y_{0},z_{0})$. After setting $\vec{R}=(x_{0},y_{0},z_{0})-(x^{\prime},y^{\prime},z^{\prime})=(x,y,z)$, the velocity becomes
\begin{equation}
\vec{v}=\frac{\Gamma}{4\pi}\int_{L\neq z_{0}}\frac{\mathrm{d}\vec{l^{\prime}}\times \vec{R}}{|\vec{R}|^{3}}=\frac{\Gamma}{4\pi}\int_{z\neq0}\vec{f}(z)\mathrm{d}z, \label{2}
\end{equation}
where $\vec{f}(z)$ is equal to
\begin{equation}
\frac{(\frac{y}{z}-\frac{\partial y}{\partial z})\vec{i}+(\frac{\partial
x}{\partial z}-\frac{x}{z})\vec{j}+(\frac{x}{z}\frac{\partial y}{\partial
z}-\frac{y}{z}\frac{\partial x}{\partial z})\vec{k}}{z|z|(1+\frac{x^{2}+y^{2}}{z^{2}})^{\frac{3}{2}}}. \label{3}
\end{equation}
Details of the calculations are given in Appendix \ref{Appendix_1}. After expanding $\vec{f}(z)$ in a Taylor series to second order at $z=0$, we obtain the polarity of the integral in the finite interval $[-l,0^{-})\cup(0^{+},l]$. Here we take as an example the $\vec{i}$ component,
\begin{widetext}
\begin{align}
\int_{-l\ne0}^lf_i\D z=\lim_{\sigma \to0^+}\ln{(\frac{l}{\sigma})}\frac{\p^2 y}{\p z^2}|_{z=0}[1+(\frac{\p y}{\p z}|_{z=0})^2+(\frac{\p x}{\p z}|_{z=0})^2]^{-\frac{3}{2}},\label{4}
\end{align}
\end{widetext}
where $l$ is any positive finite number. To make the integral finite, we need the following condition,
\begin{equation}
\frac{\partial^{2}y}{\partial z^{2}}|_{z=0}=0, \label{5}
\end{equation}
and the value of the integral value in Eq.~\eqref{4} vanishes. The same result holds for $f_{j}$. Because $z_{0}$ is chosen arbitrarily, the conclusion of Eq.~\eqref{5} is universal, which means if the integral is finite, the second-order derivative of the vortex line must be zero. If the non-zero higher-order terms are considered, it would inevitably lead to a finite second-order derivative at a certain point and a divergent integral. Therefore, we need to consider the radius of the vortex line core, which in practice exists and alters the interval of integration. Following Batchelor in regard to the application of the LIA\cite{Batchelor1970AN}, the polar integration can be dealt with by considering the effect of the radius on the denominator of Eq.~\eqref{3}. The expression for velocity eventually reduces to
\begin{widetext}
\begin{align}
\vec{v}&=\frac{\Gamma \ln{\varepsilon}}{4\pi}[-\vec{i}\frac{\p^2 y}{\p z^2}+\vec{j}\frac{\p^2 x}{\p z^2}+\vec{k}(\frac{\p^2 y}{\p z^2}\frac{\p x}{\p z}-\frac{\p^2 x}{\p z^2}\frac{\p y}{\p z})][1+(\frac{\p x}{\p z})^2+(\frac{\p y}{\p z})^2]^{-\frac{3}{2}},\label{7}
\end{align}
\end{widetext}
where $\ln{\varepsilon}$ is a parameter. Donnelly's experimental results yielded a value for $\ln \varepsilon$ in $~^{4}\mathrm{He}$ superfluid \cite{waaaa}\cite{Barenghi1983}.

Eq.~\eqref{7} shows that the self-induced velocity is only related to the first and second-order derivatives of the vortex line, which is consistent with the conclusion that the velocity is only related to the curvature of the vortex line. Note that we do not need to pay attention to the specific meaning of $x,y$ here because
\begin{equation}
\frac{\mathrm{d}^{2}x_{0}}{\mathrm{d}z_{0}^{2}}=\frac{\mathrm{d}^{2}x}{\mathrm{d}z^{2}},\text{ }\frac{\mathrm{d}^{2}y_{0}}{\mathrm{d}z_{0}^{2}}=\frac{\mathrm{d}^{2}y}{\mathrm{d}z^{2}}
\end{equation}
and the first-order derivative only involves the square term here. Therefore, in the following derivation, we do not distinguish between the position of the field point $(x_{0},y_{0},z_{0})$ and the position vector $(x,y,z)$. After resetting the position of the field point to $(x,y,z)$, the velocity can be expressed as
\begin{equation}
\vec{v}=\frac{\mathrm{d}}{\mathrm{d}t}(x,y,z).
\end{equation}

Then, by introducing a complex number,
\begin{equation}
\psi=x+\mathrm{i}y, \label{10}
\end{equation}
the equations of motion are expressed as
\begin{widetext}
\begin{align}
\left \{
\begin{aligned}
\frac{\D \psi}{\D t}&=\mathrm{i}\frac{\Gamma \ln{\varepsilon}}{4\pi}\frac{\p^2 \psi}{\p z^2}/\sqrt{1+\frac{\p \psi}{\p z}\frac{\p \psi^*}{\p z}}^3,\\
\frac{\D z}{\D t}&=-\mathrm{i}\frac{\Gamma \ln{\varepsilon}}{8\pi}(\frac{\p^2 \psi}{\p z^2}\frac{\p \psi^*}{\p z}-\frac{\p^2 \psi^*}{\p z^2}\frac{\p \psi}{\p z})/\sqrt{1+\frac{\p \psi}{\p z}\frac{\p \psi^*}{\p z}}^3.
\end{aligned}
\right.\label{11}
\end{align}
\end{widetext}
Given the relations for the derivatives, a functional expression is finally derived,
\begin{equation}
\mathrm{i}\frac{\partial \psi}{\partial t}=-\frac{\Gamma \ln{\varepsilon}}{4\pi }(\frac{\psi^{\prime}}{\sqrt{1+\psi^{\ast^{\prime}}\psi^{\prime}}})^{\prime}.
\label{12}
\end{equation}
If we introduce the linear approximation, we derive a Schr\"{o}dinger-like equation for the vortex line,
\begin{equation}
\mathrm{i}\frac{\partial \psi}{\partial t}=-\frac{\Gamma \ln{\varepsilon}}{4\pi}\frac{\partial^{2}\psi}{\partial z^{2}}.\label{13}
\end{equation}

Next, we determine a sufficient condition for the validity of the LLIA. Consider the Kelvin wave
\begin{equation}
\psi=a\mathrm{e}^{\mathrm{i}(kz-\omega t)},
\end{equation}
where $a$ denotes the radius (amplitude), $k$ the wave number, and $\omega$ the angular frequency of the Kelvin wave. As the Kelvin wave is the eigen-solution of Eqs.~\eqref{12} and \eqref{13}, the dispersion relations are
\begin{align}
\omega_{\mathrm{n}} &  =\frac{\Gamma \ln{\varepsilon}}{4\pi}\frac{k^{2}}{\sqrt{1+a^{2}k^{2}}},\\
\omega_{\mathrm{l}} &  =\frac{\Gamma \ln{\varepsilon}}{4\pi}k^{2}.
\end{align}
where $\omega_{\mathrm{n}}(k)$ establishes the dispersion relation of the non-linear equation Eq.~\eqref{12} and $\omega_{\mathrm{l}}(k)$ that for the linear equation Eq.~\eqref{13}. Clearly, if $a^{2}k^{2}\ll1$, the LLIA is valid. To make this condition more explicit, we introduce a characteristic evolution time $T_{0}$; at the beginning of the revolution process ($t=0$), the phases of the two Kelvin waves are $\varphi_{\mathrm{n}}=0$ are $\varphi_{\mathrm{l}}=0$; after time $T_{0}$, the phase difference is $\pi/2$, that is
\begin{equation}
\varphi_{\mathrm{l}}-\varphi_{\mathrm{n}}=(\omega_{\mathrm{l}}-\omega
_{\mathrm{n}})T_{0}\sim \frac{\pi}{2}.
\end{equation}
We then find the LLIA condition to be
\begin{equation}
a<\frac{2\pi \sqrt{k^{2}T_{0}\Gamma \ln{\varepsilon}-\pi^{2}}}{k(k^{2}T_{0}\Gamma\ln{\varepsilon}-2\pi^{2})}.\label{com}
\end{equation}
The physical meaning of this condition is that the effect of non-linearity may be ignored if the radius of the Kelvin wave does not go beyond the constant value determined from the algebraic expression on the right-hand side of Eq.~\eqref{com} during the revolution time of $T_{0}$.

Indeed, in the$~^{4}\mathrm{He}$ superfluid experiment, the radius of the Kelvin wave is about $10^{-2}$ to $10^{-4}\mathrm{cm}$ \cite{Baggaley2012}, the vorticity is $\Gamma\approx9.97\times10^{-8}\mathrm{m^{2}/s}$ \cite{Fonda2014Direct}, the wave-number is about $5000\mathrm{m}^{-1}$ \cite{Fonda2014Direct}, and $\ln \varepsilon \approx0.8$ \cite{waaaa}\cite{Barenghi1983}. Hence, the characteristic evolution time is about
\[
T_{0}=\frac{2\pi^{2}}{k^{2}\Gamma \ln{\varepsilon}(1-\frac{1}{\sqrt
{1+a^{2}k^{2}}})}\approx10\sim100\mathrm{s}.
\]
This implies that the evolution time may last at least for 10 seconds without breaking the LLIA in the$~^{4}\mathrm{He}$ superfluid experiment. Moreover, in the limit $a\rightarrow0$, the LLIA is well defined.

\section{Physical quantities of vortexlines}\label{sec2}
We have obtained the Schr\"{o}dinger equation based on the Biot--Savart equation under the LLIA. Next, we derive expressions for the operators representing linear momentum, angular momentum, and energy (i.e., the Hamiltonian) from their original definitions in a fluid. It is clearly shown that the operators are similar to those in QM.

\subsection*{Momentum Operator}
We change the coordinate system so that the central axis of the vortex line lies along the $z$-axis, and consider instances for which $\psi$ is relatively small,
\[
\left.  \begin{aligned} \int_{L}x\D z&=0\\ \int_{L}y\D z&=0 \end{aligned}\right \}
\rightarrow \int_{L}\psi\D z=0.
\]

The definition of the momentum of the vortex is\cite{woyundonglilun}
\begin{equation}
\vec{p}=\frac{1}{2}\int_{V}\rho \vec{r}\times \vec{\omega}\mathrm{d}V,
\label{15}
\end{equation}
where $\rho$ is the density of the fluid. The origin $o=(0,0,0)$ is the reference point for $\vec{r}$. For the vortex line, the projection of the momentum along the $z$-axis is
\begin{equation}
p_{z}=\frac{\Gamma \rho}{2}\int_{L}(x\mathrm{d}y-y\mathrm{d}x), \label{16}
\end{equation}
the details of which are given in Appendix \ref{Appendix_2}. Using the definition of $\psi$, we get
\[
p_{z}=-\mathrm{i}\frac{\Gamma \rho}{2}\int_{L}\psi^{\ast}\frac{\partial \psi
}{\partial z}\mathrm{d}z.
\]
We introduce a `normalized' wave function $\psi_{\mathrm{n}}$,
\begin{equation}
\psi_{\mathrm{n}}=\sqrt{\frac{\pi}{V}}\psi \label{n}
\end{equation}
where $V=\int_{L}\pi \psi^{\ast}\psi \mathrm{d}z$. Then, $p_{z}$ is written in the form
\begin{equation}
p_{z}=\int_{L}\psi_{\mathrm{n}}^{\ast}(-\mathrm{i}\frac{\Gamma \rho V}{2\pi
}\frac{\partial}{\partial z})\psi_{\mathrm{n}}\mathrm{d}z. \label{20}
\end{equation}
Finally, the operator representing momentum is obtained
\begin{equation}
\hat{p}=-\mathrm{i}\frac{\Gamma \rho V}{2\pi}\frac{\partial}{\partial z}.
\label{21}
\end{equation}

The momentum of the Kelvin wave is then
\[
p=\frac{\Gamma \rho}{2}a^{2}Lk,
\]
where $L$ is the length of the interval of integration along $z$. We find the effective "de Broglie relation" to be
\begin{equation}
p=\frac{\Gamma \rho V}{2\pi}k, \label{24}
\end{equation}
where $V=\pi a^{2}L$ corresponds to our definition. In addition, the commutation relation is obtained from Eq.~\eqref{21},
\begin{equation}
\lbrack \hat{z},\hat{p}]=\mathrm{i}\frac{\Gamma \rho V}{2\pi}. \label{25}
\end{equation}

\subsection*{Angular Momentum}
\begin{figure}[ptbh]
\centering
\includegraphics[width=2.5in]{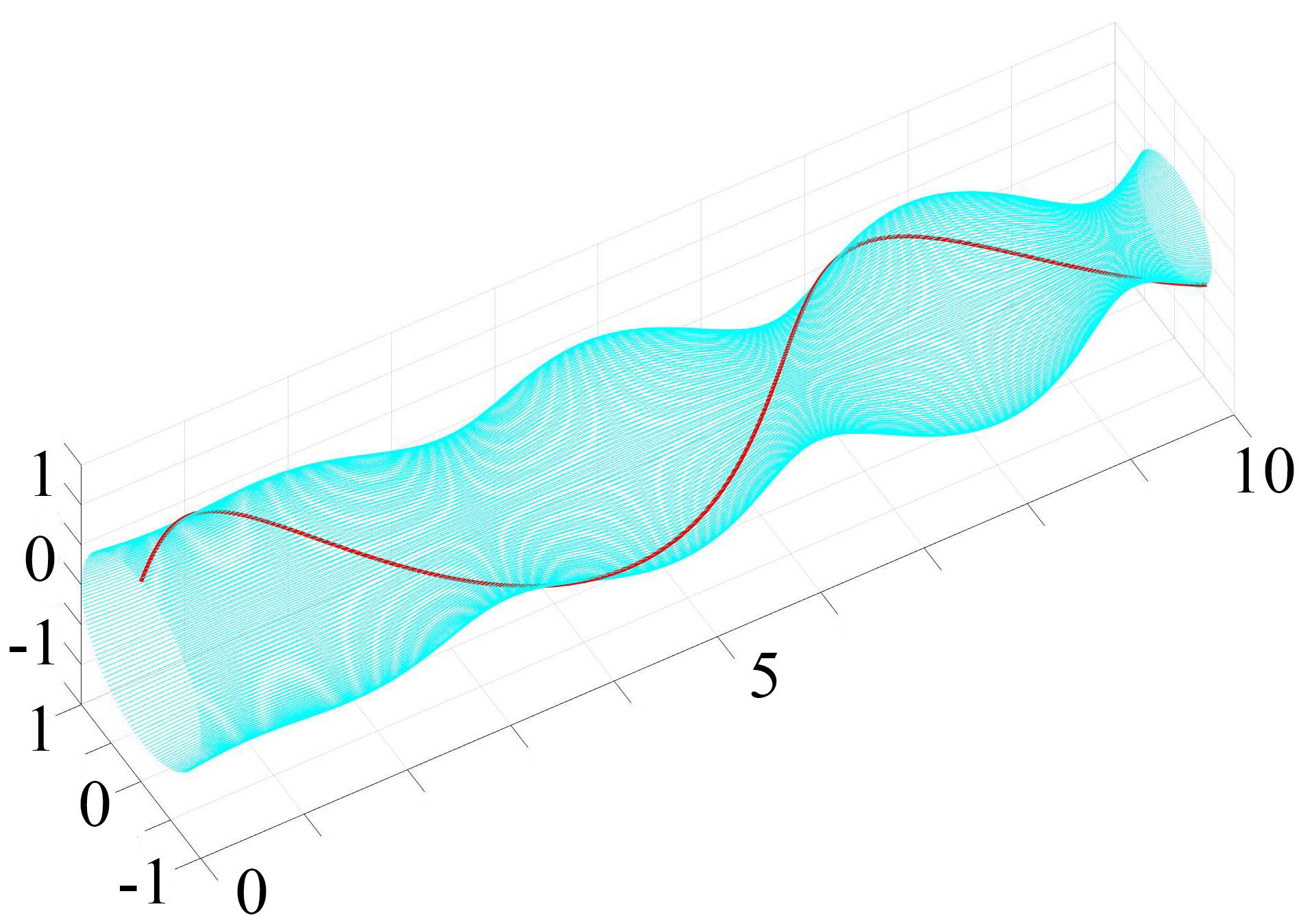}\caption{The illustration of Kelvin wave of
vortexline: the red line denotes the vortexline, and the blue one is the tube
generated by vortexline rotating around $z$-axis.}
\label{fig3:subfig}
\end{figure}

The definition for the angular momentum of a vortex is\cite{woyundonglilun}
\begin{equation}
\vec{M}=-\frac{1}{2}\int_{V}\rho r^{2}\vec{\omega}\mathrm{d}V. \label{26}
\end{equation}
By setting $(0,0,0)$ as the reference point, the projected angular momentum along the $z$-axis of the vortex line is
\[
M_{z}=-\frac{\Gamma \rho}{2}\int_{L}(x^{2}+y^{2}+z^{2})\mathrm{d}z.
\]
For the length of the vortex line, the integral is infinite. Hence we must subtract the trivial straight-vortex line to extract the additional (effective) angular momentum of the curved vortex line
\[
L_{z}=M_{z}-M_{0z}.
\]
Finally, we obtain
\begin{equation}
L_{z}=-\frac{\Gamma \rho}{2}\int_{L}\psi^{\ast}\psi \mathrm{d}z=-\frac
{\Gamma \rho V}{2\pi},\label{29}
\end{equation}
for which the effective angular momentum of vortex line is proportional to $V$, the `volume' of the vortex line. As indicated in Fig.~\ref{fig3:subfig}, when rotating the vortex line (red line) around the $z$-axis, the Kelvin wave forms a tube (blue lines), and the definition of $V$ is just the volume inside the tube. As $\Gamma$ and $\rho$ are considered as constants and $V$ can be proved to be constant because of the conservation of angular momentum, we define a new parameter
\begin{equation}
\hbar_{\mathrm{eff}}=\frac{\Gamma \rho V}{2\pi}.
\end{equation}
Given expressions Eqs.~\eqref{21}--\eqref{25},\eqref{29}, we refer to this parameter as the effective Planck constant.

In the field of superfluids, according to the quantization condition for a quantized vortex line (QV), we have\cite{Feynman1955Application}
\begin{equation}
\Gamma=\frac{h}{m},
\end{equation}
where $h$ is Planck's constant and $m$ the mass of one atom, which is the fundamental unit of the fluid. The total mass inside the vortex line volume is
\begin{equation}
M=\rho V=nm,
\end{equation}
where $n$ is the number of atoms. Hence, the effective Planck constant is
\begin{equation}
\hbar_{\mathrm{eff}}=n\hbar.
\end{equation}
These results are consistent with those in Ref.~\cite{Kou2017Kelvin}.

\subsection*{Hamiltonian Operator}
The definition of the Hamiltonian (kinetic energy actually) of the vortex line is\cite{woyundonglilun}
\begin{equation}
H=\frac{1}{8\pi}\iint \rho \frac{\vec{\omega}\cdot \vec{\omega}^{\prime}}
{R}\mathrm{d}V\mathrm{d}V^{\prime}. \label{34}
\end{equation}
For a vortex line under LIA, the energy is obtained as
\begin{equation}
H\overset{\text{LIA}}{\approx}\frac{\rho \Gamma^{2}}{8\pi}\int_{L}
2\ln{\varepsilon}\sqrt{(\frac{\partial x}{\partial z})^{2}+(\frac{\partial
y}{\partial z})^{2}+1}\mathrm{d}z,
\end{equation}
which is consistent with the results of Ref.~\cite{Boffetta2009Modeling}. The calculation is detailed in Appendix \ref{hahaha}.

To simplify the expression, we set the energy of the trivial solution of a straight vortex line to be zero and focus on the change in energy,
\begin{equation}
H=\frac{\rho \Gamma^{2}\ln \varepsilon}{8\pi}\int \frac{\partial \psi^{\ast}
}{\partial z}\frac{\partial \psi}{\partial z}\mathrm{d}z,
\end{equation}
where we consider the condition for LLIA. Using the definition of $\psi$ and performing an integration by parts, we obtain
\[
H=-\int \psi_{\mathrm{n}}^{\ast}\frac{\rho V\Gamma^{2}\ln \varepsilon}{8\pi^{2}
}\frac{\partial^{2}}{\partial z^{2}}\psi_{\mathrm{n}}\mathrm{d}z,
\]
where $\psi_{\mathrm{n}}$ has been defined as the `normalized' wave function as in Eq.~\eqref{n}. Hence, we arrive with the Hamiltonian operator
\begin{equation}
\hat{H}=-\frac{\rho V\Gamma^{2}\ln \varepsilon}{8\pi^{2}}\frac{\partial^{2}}{\partial z^{2}}. \label{38}
\end{equation}

\section{Effective Schr\"{o}dinger Equation and Path Integral}\label{esepi}
We next derive the effective Schr\"{o}dinger equation from the expression for the Hamiltonian and subsequently obtain the corresponding path-integral formula under LLIA.

If we multiply both sides of Eq.~\eqref{13} by the coefficient $\frac{\Gamma \rho V}{2\pi}$, then
\begin{equation}
\mathrm{i}\frac{\Gamma \rho V}{2\pi}\frac{\partial \psi}{\partial t}
=-\frac{\Gamma^{2}\rho V\ln{\varepsilon}}{8\pi^{2}}\frac{\partial^{2}\psi
}{\partial z^{2}},
\end{equation}
which takes the form
\begin{equation}
\mathrm{i}\hbar_{\mathrm{eff}}\frac{\partial \psi}{\partial t}=\hat{H}\psi.
\label{40}
\end{equation}
There are obvious similarities between the Biot--Savart equation for vortex lines under the LLIA and the Schr\"{o}dinger equation in QM describing a free particle.

The shape of a vortex line curve is called a state, which can be expressed as a linear superposition of different Kelvin waves. As discussed at the end of Sec.~\ref{sec1}, if the radius of the Kelvin wave vanishes in the mathematical limit, all the Kelvin waves satisfy the linear Schr\"{o}dinger equation. This is the completeness relation of Hilbert space. We borrow the Dirac symbol to express the states as
\begin{align}
\psi &  \rightarrow|\psi \rangle,\\
\psi(z)  &  =\langle z|\psi \rangle, \label{45}
\end{align}
and the completeness relation is denoted by
\begin{align}
\int|z\rangle \langle z|\mathrm{d}z  &  =1,\label{46}\\
\int|p\rangle \langle p|\mathrm{d}p  &  =1. \label{wanbei}
\end{align}
Using de Broglie relation, Eq.~\eqref{24}, we have
\begin{align}
\langle z|p\rangle &  =\frac{1}{\sqrt{2\pi \hbar_{\mathrm{eff}}}}
\mathrm{e}^{\mathrm{i}pz/\hbar_{\mathrm{eff}}},\label{fu1}\\
\langle p|z\rangle &  =\frac{1}{\sqrt{2\pi \hbar_{\mathrm{eff}}}}
\mathrm{e}^{-\mathrm{i}pz/\hbar_{\mathrm{eff}}}, \label{fu2}
\end{align}
which is to be used later.

\label{PI} To avoid a confusion between the Hamiltonian of a vortex line and that of a particle, we begin with Eq.~\eqref{13} rather than Eq.~\eqref{40}. From the form of the momentum operator, Eq.~\eqref{21}, the Schr\"{o}dinger equation is expressible as
\begin{equation}
\mathrm{i}\frac{\partial}{\partial t}|\psi \rangle=\frac{\pi \ln{\varepsilon}
}{\Gamma \rho^{2}V^{2}}\hat{p}^{2}|\psi \rangle.
\end{equation}
Defining the operator $\hat{Q}$ and propagator $U(t^{\prime},t)$ from time $t$ to $t^{\prime}$ as
\begin{align}
\hat{Q}  &  =\frac{\pi \ln{\varepsilon}}{\Gamma \rho^{2}V^{2}}\hat{p}^{2},\\
|\psi(t^{\prime})\rangle &  =U(t^{\prime},t)|\psi(t)\rangle,
\end{align}
respectively, and considering $\hat{Q}$ is time-independent, the propagator is written
\begin{equation}
U(t^{\prime},t)=\mathrm{e}^{-\mathrm{i}\hat{Q}(t^{\prime}-t)}.
\end{equation}

In the coordinate representation, we make use of Eqs.~\eqref{45} and \eqref{46} to obtain
\begin{equation}
\psi(z^{\prime},t^{\prime})=\int \mathrm{d}zU(z^{\prime},t^{\prime}
;z,t)\psi(z,t),
\end{equation}
where $U(z^{\prime},t^{\prime};z,t)=\langle z^{\prime}|U(t^{\prime},t)|z\rangle$ is a `matrix element' of the propagator, which we abbreviate to $U_{z}$.

We then divide $t^{\prime}-t$ into $N$ equal parts, $t^{\prime}-t=N\Delta t$, and write $t_{k}=t+k\Delta t,$ $k=0,$ $1,$ $2,\cdots$. $U_{z}$ may be expressed as a multiplication of propagators from time $t_{k}$ to $t_{k+1}$. The $k$-th matrix element is
\[
U_{zk}=\langle z_{k+1}|\mathrm{e}^{-\mathrm{i}\hat{Q}\Delta t}|z_{k}\rangle.
\]
Applying a Fourier transformation, we obtain in the limit $N\rightarrow \infty$ and $\Delta t\rightarrow0$ an expression for the matrix element,
\[
U_{zk}=\int \mathrm{d}p_{k}\frac{1}{2\pi \hbar_{\mathrm{eff}}}\mathrm{e}
^{\mathrm{i}\frac{\Delta t}{\hbar_{\mathrm{eff}}}(p_{k}\dot{z}_{k}-\frac
{\ln{\varepsilon}}{2\rho V}p_{k}^{2})}.
\]
We then have
\begin{equation}
U_{z}=\int \mathscr{D}_{\Omega}\exp[\mathrm{i}\int_{t}^{t^{\prime}}
\mathrm{d}t\frac{1}{\hbar_{\mathrm{eff}}}(p\dot{z}-\frac{\ln{\varepsilon}
}{2\rho V}p^{2})],
\end{equation}
where
\begin{equation}
\int \mathscr{D}_{\Omega}=\lim_{N\rightarrow \infty}\int \frac{\mathrm{d}
p_{N}\dots \mathrm{d}p_{1}\mathrm{d}z_{N-1}\dots \mathrm{d}z_{1}}{(2\pi
\hbar_{\mathrm{eff}})^{N}}.
\end{equation}
If we define $H^{\prime}=\frac{\ln{\varepsilon}}{2\rho V}p^{2}$ to be the Hamiltonian of a `free particle', the system satisfies the Feynman path integral. The effective mass of a free particle then becomes
\begin{equation}
m_{\mathrm{eff}}=\frac{\rho V}{\ln{\varepsilon}}\overset{\mathrm{QV}}{=}n\frac{m}{\ln{\varepsilon}}.
\end{equation}
Indeed, the Hamiltonian of the vortex line Eq.~\eqref{38} itself simplifies to
\[
\hat{H}=\frac{\hat{p}^{2}}{2m_{\mathrm{eff}}}.
\]

Finally, we describe the movement of the wave by evaluating the Feynman path integral of the particle. The Feynman propagator is expressed as
\begin{equation}
U_{z}=\int \mathscr{D}_{\Omega}\exp[\mathrm{i}\int_{t}^{t^{\prime}}
\mathrm{d}t\frac{\mathscr{L}}{\hbar_{\mathrm{eff}}}]=\int \mathscr{D}_{\Omega
}\mathrm{e}^{\mathrm{i}{S}/\hbar_{\mathrm{eff}}},
\end{equation}
where $\mathscr{L}$ is the Lagrangian of the system, and $S$ is the action.

\section{Conclusions}
\label{sec4} The motion of a vortex line is mapped onto the movement of a particle in QM under the LLIA. On the basis of the conservation of the vortex-line volume, we define an effective Planck constant, which bridges classical mechanics with QM, including the linear and angular momentum operators, as well as the Hamiltonian form of the Schr\"{o}dinger equation.

\section*{Aknowledgement}

We thank Prof. Wenan Guo and Prof. Bin Zhou for the valuable advice offered during the discussion in Beijing Normal University. This work is supported by NSFC under the grant No. 1167402. 

\begin{widetext}
\begin{appendices}
\section{Schr\"odinger~Equation}

\label{Appendix_1} \noindent The detailed calculations of Eq.\eqref{2} is
given by
\begin{align}
\vec{v}  &  =\frac{\Gamma}{4\pi}\int_{L\neq z_{0}}\frac{\mathrm{d}
\vec{l^{\prime}}\times \vec{R}}{|\vec{R}|^{3}}\\
&  =\frac{\Gamma}{4\pi}\int_{z\neq0}\frac{(y\mathrm{d} z-z\mathrm{d} y)\vec
{i}+(z\mathrm{d} x-x\mathrm{d} z)\vec{j}+(x\mathrm{d} y-y\mathrm{d} x)\vec{k}
}{\sqrt{x^{2}+y^{2}+z^{2}}^{3}}\\
&  =\frac{\Gamma}{4\pi}\int_{z\neq0}\frac{(y-z\frac{\partial y}{\partial
z})\vec{i}+(z\frac{\partial x}{\partial z}-x)\vec{j}+(x\frac{\partial
y}{\partial z}-y\frac{\partial x}{\partial z})\vec{k}}{\sqrt{x^{2}+y^{2}
+z^{2}}^{3}}\mathrm{d} z\\
&  =\frac{\Gamma}{4\pi}\int_{z\neq0}\frac{(y-z\frac{\partial y}{\partial
z})\vec{i}+(z\frac{\partial x}{\partial z}-x)\vec{j}+(x\frac{\partial
y}{\partial z}-y\frac{\partial x}{\partial z})\vec{k}}{z^{2}|z|}(1+\frac
{x^{2}+y^{2}}{z^{2}})^{-\frac{3}{2}}\mathrm{d} z\\
&  =\frac{\Gamma}{4\pi}\int_{z\neq0}\frac{(\frac{y}{z}-\frac{\partial
y}{\partial z})\vec{i}+(\frac{\partial x}{\partial z}-\frac{x}{z})\vec
{j}+(\frac{x}{z}\frac{\partial y}{\partial z}-\frac{y}{z}\frac{\partial
x}{\partial z})\vec{k}}{z|z|}(1+\frac{x^{2}+y^{2}}{z^{2}})^{-\frac{3}{2}
}\mathrm{d} z.
\end{align}

\noindent From Eq.\eqref{3} to Eq.\eqref{4}: Here we use $\vec{i}$ component
as an example to demonstrate that the integral polarity is inevitable. The
Taylor expansion of numerator is
\begin{align}
&  \left \{
\begin{aligned} &x(z)=x(0)+\frac{\p x}{\p z}|_{z=0}(z-0)+\frac{1}{2}\frac{\p^2 x}{\p z^2}|_{z=0}(z-0)^2+\frac{1}{6}\frac{\p^3 x}{\p z^3}|_{z=0}(z-0)^3+o(z^4),\\ &y(z)=y(0)+\frac{\p y}{\p z}|_{z=0}(z-0)+\frac{1}{2}\frac{\p^2 y}{\p z^2}|_{z=0}(z-0)^2+\frac{1}{6}\frac{\p^3 y}{\p z^3}|_{z=0}(z-0)^3+o(z^4), \end{aligned} \right.
\\
\to &  \left \{
\begin{aligned} &\frac{x}{z}=\frac{\p x}{\p z}|_{z=0}+\frac{1}{2}\frac{\p^2 x}{\p z^2}|_{z=0}z+\frac{1}{6}\frac{\p^3 x}{\p z^3}|_{z=0}z^2+o(z^3),\\ &\frac{y}{z}=\frac{\p y}{\p z}|_{z=0}+\frac{1}{2}\frac{\p^2 y}{\p z^2}|_{z=0}z+\frac{1}{6}\frac{\p^3 y}{\p z^3}|_{z=0}z^2+o(z^3), \end{aligned} \right.
\end{align}
it is necessary to consider the definition of $x,y,z$, when $z\to0$, $x,y\to0$.

The $\vec{i}$ component of $\vec{f}(z)$ can be expressed as
\begin{align}
f_{i}(z)  &  =[\frac{1}{2}\frac{\partial^{2} y}{\partial z^{2}}|_{z=0}
z+\frac{1}{6}\frac{\partial^{3} y}{\partial z^{3}}|_{z=0}z^{2}+o(z^{3}
)]\frac{1}{z|z|}[1+(\frac{\partial y}{\partial z}|_{z=0})^{2}+(\frac{\partial
x}{\partial z}|_{z=0})^{2}+o(z^{3})]^{-\frac{3}{2}}\\
&  =\left \{
\begin{aligned} &[\frac{1}{2}\frac{\p^2 y}{\p z^2}|_{z=0}\frac{1}{z}+\frac{1}{6}\frac{\p^3 y}{\p z^3}|_{z=0}][1+(\frac{\p y}{\p z}|_{z=0})^2+(\frac{\p x}{\p z}|_{z=0})^2]^{-\frac{3}{2}},z>0,\\ &-[\frac{1}{2}\frac{\p^2 y}{\p z^2}|_{z=0}\frac{1}{z}+\frac{1}{6}\frac{\p^3 y}{\p z^3}|_{z=0}][1+(\frac{\p y}{\p z}|_{z=0})^2+(\frac{\p x}{\p z}|_{z=0})^2]^{-\frac{3}{2}},z<0. \end{aligned} \right.
\end{align}
So the integral is
\begin{align}
\int_{-l\ne0}^{l}f_{i}\mathrm{d} z=  &  \lim_{\sigma \to0^{+}}\int
_{-l}^{-\sigma}-[\frac{1}{2}\frac{\partial^{2} y}{\partial z^{2}}|_{z=0}
\frac{1}{z}+\frac{1}{6}\frac{\partial^{3} y}{\partial z^{3}}|_{z=0}
][1+(\frac{\partial y}{\partial z}|_{z=0})^{2}+(\frac{\partial x}{\partial
z}|_{z=0})^{2}]^{-\frac{3}{2}}\mathrm{d} z\nonumber \\
&  +\int_{\sigma}^{l}[\frac{1}{2}\frac{\partial^{2} y}{\partial z^{2}}
|_{z=0}\frac{1}{z}+\frac{1}{6}\frac{\partial^{3} y}{\partial z^{3}}
|_{z=0}][1+(\frac{\partial y}{\partial z}|_{z=0})^{2}+(\frac{\partial
x}{\partial z}|_{z=0})^{2}]^{-\frac{3}{2}}\mathrm{d} z\\
=  &  \lim_{\sigma \to0^{+}}[\ln{(\frac{l}{\sigma})}\frac{\partial^{2}
y}{\partial z^{2}}|_{z=0}+0][1+(\frac{\partial y}{\partial z}|_{z=0}
)^{2}+(\frac{\partial x}{\partial z}|_{z=0})^{2}]^{-\frac{3}{2}},
\end{align}
where $l$ can be any finite number.

We next show how to derive Eq.\eqref{11}. The equations of motion are
\[
\left \{
\begin{aligned} \dot{\psi}=\frac{\D \psi}{\D t}&=\mathrm{i}\frac{\Gamma \ln{\varepsilon}}{4\pi}\frac{\p^2 \psi}{\p z^2}/\sqrt{1+\frac{\p \psi}{\p z}\frac{\p \psi^*}{\p z}}^3,\\ \dot{z}=\frac{\D z}{\D t}&=-\mathrm{i}\frac{\Gamma \ln{\varepsilon}}{8\pi}(\frac{\p^2 \psi}{\p z^2}\frac{\p \psi^*}{\p z}-\frac{\p^2 \psi^*}{\p z^2}\frac{\p \psi}{\p z})/\sqrt{1+\frac{\p \psi}{\p z}\frac{\p \psi^*}{\p z}}^3. \end{aligned}\right.
\]
Considering the relation of derivative symbols, the equation is
\begin{align}
\frac{\partial \psi}{\partial t}  &  =\frac{\mathrm{d}\psi}{\mathrm{d}t}
-\frac{\mathrm{d}z}{\mathrm{d}t}\frac{\partial \psi}{\partial z}\\
&  =\frac{\mathrm{i}\Gamma \ln{\varepsilon}}{4\pi}[\frac{\partial^{2}\psi
}{\partial z^{2}}+\frac{1}{2}(\frac{\partial^{2}\psi}{\partial z^{2}}
\frac{\partial \psi^{\ast}}{\partial z}-\frac{\partial^{2}\psi^{\ast}}{\partial
z^{2}}\frac{\partial \psi}{\partial z})\frac{\partial \psi}{\partial z}
]/\sqrt{1+\frac{\partial \psi}{\partial z}\frac{\partial \psi^{\ast}}{\partial
z}}^{3}\\
&  =\frac{\mathrm{i}\Gamma \ln{\varepsilon}}{4\pi}[(\frac{\partial^{2}\psi
}{\partial z^{2}}+\frac{\partial^{2}\psi}{\partial z^{2}}\frac{\partial
\psi^{\ast}}{\partial z}\frac{\partial \psi}{\partial z})-\frac{1}{2}
(\frac{\partial^{2}\psi}{\partial z^{2}}\frac{\partial \psi^{\ast}}{\partial
z}+\frac{\partial^{2}\psi^{\ast}}{\partial z^{2}}\frac{\partial \psi}{\partial
z})\frac{\partial \psi}{\partial z}]/\sqrt{1+\frac{\partial \psi}{\partial
z}\frac{\partial \psi^{\ast}}{\partial z}}^{3}\\
&  =\frac{\mathrm{i}\Gamma \ln{\varepsilon}}{4\pi}[\frac{\partial^{2}\psi
}{\partial z^{2}}(1+\frac{\partial \psi^{\ast}}{\partial z}\frac{\partial \psi
}{\partial z})-\frac{1}{2}(\frac{\partial^{2}\psi}{\partial z^{2}}
\frac{\partial \psi^{\ast}}{\partial z}+\frac{\partial^{2}\psi^{\ast}}{\partial
z^{2}}\frac{\partial \psi}{\partial z})\frac{\partial \psi}{\partial z}
]/\sqrt{1+\frac{\partial \psi}{\partial z}\frac{\partial \psi^{\ast}}{\partial
z}}^{3}\\
&  =\frac{\mathrm{i}\Gamma \ln{\varepsilon}}{4\pi}[\frac{\partial^{2}\psi
}{\partial z^{2}}\sqrt{1+\frac{\partial \psi^{\ast}}{\partial z}\frac
{\partial \psi}{\partial z}}-\frac{\frac{\partial^{2}\psi}{\partial z^{2}}
\frac{\partial \psi^{\ast}}{\partial z}+\frac{\partial^{2}\psi^{\ast}}{\partial
z^{2}}\frac{\partial \psi}{\partial z}}{2\sqrt{1+\frac{\partial \psi^{\ast}
}{\partial z}\frac{\partial \psi}{\partial z}}}\frac{\partial \psi}{\partial
z}]/\sqrt{1+\frac{\partial \psi}{\partial z}\frac{\partial \psi^{\ast}}{\partial
z}}^{2}\\
&  =\frac{\mathrm{i}\Gamma \ln{\varepsilon}}{4\pi}[(\frac{\partial}{\partial
z}\frac{\partial \psi}{\partial z})\sqrt{1+\frac{\partial \psi^{\ast}}{\partial
z}\frac{\partial \psi}{\partial z}}-(\frac{\partial}{\partial z}\sqrt
{1+\frac{\partial \psi^{\ast}}{\partial z}\frac{\partial \psi}{\partial z}
})\frac{\partial \psi}{\partial z}]/\sqrt{1+\frac{\partial \psi}{\partial
z}\frac{\partial \psi^{\ast}}{\partial z}}^{2}\\
&  =\frac{\mathrm{i}\Gamma \ln{\varepsilon}}{4\pi}\frac{\partial}{\partial
z}\frac{\frac{\partial \psi}{\partial z}}{\sqrt{1+\frac{\partial \psi^{\ast}
}{\partial z}\frac{\partial \psi}{\partial z}}}.
\end{align}
By rewriting the equation in another form, we derive Eq.\eqref{12}
\begin{equation}
\mathrm{i}\frac{\partial \psi}{\partial t}=-\frac{\Gamma \ln{\varepsilon}}{4\pi
}(\frac{\psi^{\prime}}{\sqrt{1+\psi^{\ast^{\prime}}\psi^{\prime}}})^{\prime}.
\end{equation}

\section{Operators}

\label{Appendix_2}

\subsection{Momentum}

We show the detailed calculations of momentum operator. The momentum is
defined by
\begin{equation}
\vec{p}=\frac{1}{2}\int_{V}\rho \vec{r}\times \vec{\omega}\mathrm{d}V.
\end{equation}
For vortexline, we have
\[
\vec{p}=\frac{\Gamma \rho}{2}\int_{L}\vec{r}\times \mathrm{d}\vec{l}
=\frac{\Gamma \rho}{2}\int_{L}(y\mathrm{d}z-z\mathrm{d}y)\vec{i}+(z\mathrm{d}
x-x\mathrm{d}z)\vec{j}+(x\mathrm{d}y-y\mathrm{d}x)\vec{k},
\]
where we've already set the reference point of $\vec{r}$ at $o=(0,0,0)$. The
momentum of $z$-axis is
\begin{align}
p_{z}=\frac{\Gamma \rho}{2}\int_{L}(x\mathrm{d}y-y\mathrm{d}x)  &
=\frac{\Gamma \rho}{2}\int_{L}\frac{\psi+\psi^{\ast}}{2}\frac{\partial
(\psi-\psi^{\ast})}{2\mathrm{i}\partial z}-\frac{\psi-\psi^{\ast}}
{2\mathrm{i}}\frac{\partial(\psi+\psi^{\ast})}{2\partial z}\mathrm{d}z\\
&  =-\mathrm{i}\frac{\Gamma \rho}{4}(\int_{L}\psi^{\ast}\frac{\partial \psi
}{\partial z}\mathrm{d}z-\int_{L}\psi \frac{\partial \psi^{\ast}}{\partial
z}\mathrm{d}z)\\
&  =-\mathrm{i}\frac{\Gamma \rho}{4}(\int_{L}\psi^{\ast}\frac{\partial \psi
}{\partial z}\mathrm{d}z-\psi \psi^{\ast}|_{-\infty}^{+\infty}+\int_{L}
\psi^{\ast}\frac{\partial \psi}{\partial z}\mathrm{d}z)\\
&  =-\mathrm{i}\frac{\Gamma \rho}{2}\int_{L}\psi^{\ast}\frac{\partial \psi
}{\partial z}\mathrm{d}z+\mathrm{i}\frac{\Gamma \rho}{4}\psi \psi^{\ast
}|_{-\infty}^{+\infty}.
\end{align}
When $\psi$ is a kind of `bound state', there is $\psi \psi^{\ast}|_{-\infty
}^{+\infty}=0$.

\subsection{Angular Momentum}

The augular momentum is defined by
\begin{equation}
\vec{M}=-\frac{1}{2}\int_{V}\rho r^{2}\vec{\omega}\mathrm{d} V.
\end{equation}
As to vortexline, we have
\[
\vec{M}=-\frac{\Gamma \rho}{2}\int_{L}r^{2}\mathrm{d} \vec{l}=-\frac{\Gamma
\rho}{2}\int_{L}(x^{2}+y^{2}+z^{2})(\vec{i}\mathrm{d} x+\vec{j}\mathrm{d}
y+\vec{k}\mathrm{d} z),
\]
where $(0,0,0)$ is the reference point. And the $z$-axis component is
\begin{equation}
M_{z}=-\frac{\Gamma \rho}{2}\int_{L}(x^{2}+y^{2}+z^{2})\mathrm{d} z.
\end{equation}

The projected angular momentum is given by
\begin{align}
L_{z}  &  =M_{z}-M_{0z}\\
&  =-\frac{\Gamma \rho}{2}\int_{L}(x^{2}+y^{2}+z^{2}-z^{2})\mathrm{d}z\\
&  =-\frac{\Gamma \rho}{2}\int_{L}(x^{2}+y^{2})\mathrm{d}z\\
&  =-\frac{\Gamma \rho}{2}\int_{L}\psi^{\ast}\psi \mathrm{d}z=-\frac{\Gamma \rho
V}{2\pi},
\end{align}
where we have assumed that $\Gamma$ is along the positive direction of $z$-axis.

\subsection{Hamiltonian}

\label{hahaha}

Finally, we calculate the Hamiltonian,
\begin{equation}
H=\frac{1}{8\pi}\iint \rho \frac{\vec{\omega}\cdot \vec{\omega}^{\prime}}
{R}\mathrm{d}V\mathrm{d}V^{\prime}.
\end{equation}
As to vortexline, the expression is
\begin{align}
H  &  =\frac{\rho \Gamma^{2}}{8\pi}\iint \frac{\mathrm{d}\vec{l}\cdot
\mathrm{d}\vec{l}^{\prime}}{R}\\
&  =\frac{\rho \Gamma^{2}}{8\pi}\int_{L}\mathrm{d}z\int_{L^{\prime}\neq
z}\mathrm{d}z^{\prime}\frac{\frac{\partial x}{\partial z}\frac{\partial
x^{\prime}}{\partial z}+\frac{\partial y}{\partial z}\frac{\partial y^{\prime
}}{\partial z}+1}{\sqrt{(x-x^{\prime2}+(y-y^{\prime2}+(z-z^{\prime2}}}\\
&  \overset{\text{LIA}}{\approx}\frac{\rho \Gamma^{2}}{8\pi}\int_{L}
\mathrm{d}z\int_{z-l\neq z0}^{z+l}\mathrm{d}z^{\prime}\frac{\frac{\partial
x}{\partial z}\frac{\partial x}{\partial z}+\frac{\partial y}{\partial z}
\frac{\partial y}{\partial z}+1}{\sqrt{[(\frac{\partial x}{\partial z}
)^{2}+(\frac{\partial y}{\partial z})^{2}+1][(z-z^{\prime2}+\sigma^{2}]}}.
\end{align}
we utilize a method under LIA for processing velocity integral polarity to
derive integral value near $z^{\prime}\rightarrow z$. The main contribution of
the integral exists near the area of polarity. Besides, we further approximate
coordinate differences to linear items of Taylor Expansion, $\frac
{x-x^{\prime}}{z-z^{\prime}}\rightarrow \frac{\partial x}{\partial z}
,\frac{y-y^{\prime}}{z-z^{\prime}}\rightarrow \frac{\partial y}{\partial z}$,
and the first order derivatives are equal in the integral interval
($\frac{\partial x^{\prime}}{\partial z}\rightarrow \frac{\partial x}{\partial
z},\frac{\partial y^{\prime}}{\partial z}\rightarrow \frac{\partial y}{\partial
z}$). So, the expression goes to
\begin{align}
H  &  \overset{\text{LIA}}{\approx}\frac{\rho \Gamma^{2}}{8\pi}\int_{L}
2\ln{\varepsilon}\sqrt{(\frac{\partial x}{\partial z})^{2}+(\frac{\partial
y}{\partial z})^{2}+1}\mathrm{d}z\\
&  =\frac{\rho \Gamma^{2}\ln{\varepsilon}}{4\pi}\int_{L}\sqrt{1+\frac
{\partial \psi^{\ast}}{\partial z}\frac{\partial \psi}{\partial z}}\mathrm{d}z\\
&  \overset{\text{LLIA}}{\approx} \frac{\rho \Gamma^{2}\ln{\varepsilon}}{4\pi
}\int_{L}(1+\frac{1}{2}\frac{\partial \psi^{\ast}}{\partial z}\frac
{\partial \psi}{\partial z})\mathrm{d}z.
\end{align}

Finally, we have
\begin{align}
H  &  =\frac{\rho \Gamma^{2}\ln \varepsilon}{8\pi}\int \frac{\partial \psi^{\ast}
}{\partial z}\frac{\partial \psi}{\partial z}\mathrm{d} z\\
&  =\frac{\rho \Gamma^{2}\ln \varepsilon}{8\pi}(\frac{\partial \psi}{\partial
z}\psi^{\ast}|_{-\infty}^{\infty}-\int \psi^{\ast}\frac{\partial^{2}\psi
}{\partial z^{2}}\mathrm{d} z)\\
&  =-\int \psi^{\ast}\frac{\rho \Gamma^{2}\ln \varepsilon}{8\pi}\frac
{\partial^{2}}{\partial z^{2}}\psi \mathrm{d} z\\
&  =-\int \psi_{\mathrm{n}}^{\ast}\frac{\rho V\Gamma^{2}\ln \varepsilon}
{8\pi^{2}}\frac{\partial^{2}}{\partial z^{2}}\psi_{\mathrm{n}}\mathrm{d} z
\end{align}
\end{appendices}
\end{widetext}
\bibliographystyle{ieeetr}
\bibliography{EQMref}

\begin{thebibliography}{10}

\bibitem{kelvin}
W.~Thomson, ``{On Vortex Motion},'' {\em Transactions of the Royal Society of
  Edinburgh}, vol.~25, pp.~217--260, 1867.

\bibitem{ajhj}
H.E.Hall {\em Proc.Roy.Soc.A}, vol.~245, p.~546, 1958.

\bibitem{Batchelor1970AN}
G.~K. Batchelor, {\em An Introduction to Fluid Dynamics}.
\newblock At the University Press, 1970.

\bibitem{1972JFM....51..477H}
H.~{Hasimoto}, ``{A soliton on a vortex filament},'' {\em Journal of Fluid
  Mechanics}, vol.~51, pp.~477--485, 1972.

\bibitem{waaaa}
C.~R.J.Donnelly, ``{The observed properties of liquid helium at the saturated
  vapor pressure},'' {\em J.Phys.Chem.Ref.Data}, vol.~27, pp.~1217--1274, 1998.

\bibitem{Boffetta2009Modeling}
Boffetta.G., Celani.A., Dezzani.D., Laurie.J., and Nazarenko.S., ``Modeling
  kelvin wave cascades in superfluid helium,'' {\em Journal of Low Temperature
  Physics}, vol.~156, no.~3-6, pp.~193--214, 2009.

\bibitem{Salman2014Multiple}
Salman.H, ``Multiple breathers on a vortex filament,'' p.~012005, 2014.

\bibitem{Salman2013Breathers}
H.~Salman, ``Breathers on quantized superfluid vortices,'' {\em Physical Review
  Letters}, vol.~111, no.~16, p.~165301, 2013.

\bibitem{Vinen2002}
W.~F. Vinen and J.~J. Niemela, ``Quantum turbulence,'' {\em Journal of Low
  Temperature Physics}, vol.~128, pp.~167--231, Sep 2002.

\bibitem{Laurie2009The}
J.~Laurie, V.~S. L'Vov, S.~Nazarenko, and O.~Rudenko, ``The interaction of
  kelvin waves and the non-locality of the energy transfer in superfluids,''
  {\em Physical Review B}, vol.~81, no.~10, p.~104526, 2009.

\bibitem{1926AnP...384..361S}
E.~{Schr{\"o}dinger}, ``{Quantisierung als Eigenwertproblem},'' {\em Annalen
  der Physik}, vol.~384, pp.~361--376, 1926.

\bibitem{333}
B.~Malomed, {\em Nonlinear \schr~ Equations}.
\newblock Encyclopedia of Nonlinear Science, 2005.

\bibitem{Batygin2018Schr}
K.~Batygin, ``\schr evolution of self-gravitating discs,'' {\em Monthly Notices
  of the Royal Astronomical Society}, vol.~475, 2018.

\bibitem{Barenghi1983}
C.~F. Barenghi, R.~J. Donnelly, and W.~F. Vinen, ``Friction on quantized
  vortices in helium ii. a review,'' {\em Journal of Low Temperature Physics},
  vol.~52, pp.~189--247, Aug 1983.

\bibitem{Baggaley2012}
A.~W. Baggaley, ``The sensitivity of the vortex filament method to different
  reconnection models,'' {\em Journal of Low Temperature Physics}, vol.~168,
  pp.~18--30, Jul 2012.

\bibitem{Fonda2014Direct}
E.~Fonda, D.~P. Meichle, N.~T. Ouellette, S.~Hormoz, and D.~P. Lathrop,
  ``Direct observation of kelvin waves excited by quantized vortex
  reconnection,'' {\em Proc Natl Acad Sci U S A}, vol.~111 Suppl 1,
  pp.~4707--4710, 2014.

\bibitem{woyundonglilun}
B.~Tong, X.~Yin, and K.~Zhu, {\em Theory of Vorticity}.
\newblock China Science and Technology University Press, 2009.

\bibitem{Feynman1955Application}
R.~P. Feynman, ``Application of quantum mechanics to liquid helium,'' {\em
  Helium}, pp.~268--313, 1955.

\bibitem{Kou2017Kelvin}
S.~P. Kou, ``Kelvin wave and knot dynamics on entangled vortices,'' {\em
  International Journal of Modern Physics B}, p.~1750241, 2017.

\end{thebibliography}

\end{document}